\documentclass[preprint,preprintnumbers,amsmath,amssymb,aps,prl]{revtex4}

\usepackage{graphicx}
\usepackage{dcolumn}
\usepackage{bm}

\begin{document}

\title{An Exact Model of Fluctuations in Gene Expression}
\author{William W. Chen}
\thanks{These authors contributed equally to this work.}
\affiliation{%
Harvard University, Department of Biophysics, 240 Longwood Avenue, Boston Massachusetts 02115
}%
\author{Jeremy L. England}
\thanks{These authors contributed equally to this work.}
\affiliation{%
}
\author{Eugene I. Shakhnovich}%
 \email{eugene@belok.harvard.edu}
\affiliation{%
Harvard University, Department of Chemistry and Chemical Biology, 12 Oxford Street, Cambridge, Massachusetts 02138
}%


\date{\today}

\begin{abstract}
Fluctuations in the measured mRNA levels of unperturbed cells under fixed conditions have often been viewed as an
impediment to the extraction of information from expression profiles. Here, we argue that such expression fluctuations
should themselves be studied as a source of valuable information about the underlying dynamics of genetic networks. By
analyzing microarray data taken from Saccharomyces cerevisiae, we demonstrate that correlations in expression
fluctuations have a highly statistically significant dependence on gene function, and furthermore exhibit a remarkable
scale-free network structure. We therefore present what we view to be the simplest phenomenological model of a genetic
network which can account for the presence of biological information in transcript level fluctuations. We proceed to
exactly solve this model using a path integral technique and derive several quantitative predictions. Finally, we
propose several experiments by which these predictions might be rigorously tested. 
\end{abstract}

\maketitle
The classical approach to molecular biology is to characterize living
systems principally by analyzing their individual components and determining how these
components interact with each other on a case-by-case basis.  In contrast, the
recent development of high-throughput gene expression assays has
made it possible to study the cell's web of
interacting gene products as a single, complex system with many degrees of
freedom.~\cite{schena_brown,wen_somogyi}.  

This approach has proved to be
useful not only as a new tool for medical diagnosis~\cite{veer_friend, alizadeh_staudt},
but also as a fount of novel insights into how biological systems function as a
whole~\cite{arava_herschlag, laub_shapiro, spellman_futcher}.

It has previously been observed in the case of gene microarray assays that data which are taken from
repetitions of the same experiment can nonetheless exhibit a substantial amount of variation.
These fluctuations in measured gene expression have typically been treated as a drawback to using microarrays, 
so much so that even where it has been observed that expression fluctuations
do contain information about gene function, efforts still have been focused on finding ways to deconvolute their impact 
from the data~\cite{hughes}.  
This attitude is mirrored in the characteristics of many previously proposed theoretical models of gene networks;  despite 
ample evidence that gene expression is a fundamentally ``noisy,'' stochastic process~\cite{thattai_oudenaarden, hasty_collins, elowitz_swain},
many such models have attempted
to describe whole-network dynamics using deterministic differential equations or Boolean models~\cite{collins_gardner, kauffman_troein}

In this work, we present analyses of yeast microarray data which illustrate the value of expression fluctuations as a potential source
of information about genetic networks.  We therefore propose what we
believe to be the simplest analytical model of a whole genetic network which can explain the biological information
we observe in these fluctuations.  
In what follows, we first solve the model using a path integral technique and derive from it several quantitative predictions.  
Subsequently, we propose experiments by which these predictions could be straightforwardly tested.

\section{Biological Information in Expression Fluctuations}

In their large scale study of yeast expression profiles conducted in 2000, Hughes et al. ~\cite{hughes} 
carried out 63 identically prepared gene expression assays of 
yeast cells under normal conditions. 
Though the investigators noted in passing that correlations in the 
fluctuations of their control data could be used to cluster genes by function, 
their aim was principally to gauge the magnitude of this ``intrinsic biological noise'' 
in order to be able subtract its effects from their subsequent experiments.  
Here, we present similar analyses of the same set of control data, 
arguing that the amount of biological information contained in such noise 
merits further study in its own right.  

Using data from the same 63 assays, we constructed a series of 
gene networks.  Networks were made by representing 
each gene as a node in a graph, and by drawing links 
between gene pairs when the Pearson correlations of their levels 
of expression exceeded some cut-off threshhold in magnitude.  
The Pearson correlation between the expression
levels of two genes is defined as follows:
\begin{equation}
\frac{\langle x y \rangle - \langle x \rangle \langle y \rangle}
{\sqrt{\langle x^2 \rangle - \langle x \rangle^2} \,
\sqrt{\langle y^2 \rangle - \langle y \rangle^2}}
\end{equation}
where $x$ and $y$ are the expression levels of the genes.
The averaging is taken over the 63 experiments. 
The Pearson correlation measures how the natural fluctuations 
of any two genes track each other in identically prepared 
experiments.  Two types of analyses were applied to these 
correlation networks.

In a fashion very similar to the original analysis done by 
Hughes et al., we gauged the ability of these networks to assign genes 
with identical functional classifications~\cite{mips} to the same cluster.  For each 
constructed network, clusters were defined as sets of genes belonging to 
disconnected subgraphs.  
By tuning the Pearson correlation cut-off threshhold, we constructed networks with
varying ``connectedness'' and thus varying sizes of the giant component.  
The giant component is defined as the largest disconnected subgraph
of the network.  At one extreme, when the cut-off threshhold is 
very low, the network is completely connected and the giant component 
is the entire set of genes.  At the other extreme, when
the cut-off threshhold is very high, the network is completely disconnected
and all subgraphs are orphan nodes.  Thus the size of the
giant component varies between 1 and $N$ (where $N$ is the number of genes)
and serves as a natural scale for the ``connectedness'' of the network.

\begin{figure}[t]
\resizebox{\columnwidth}{!}{
\includegraphics{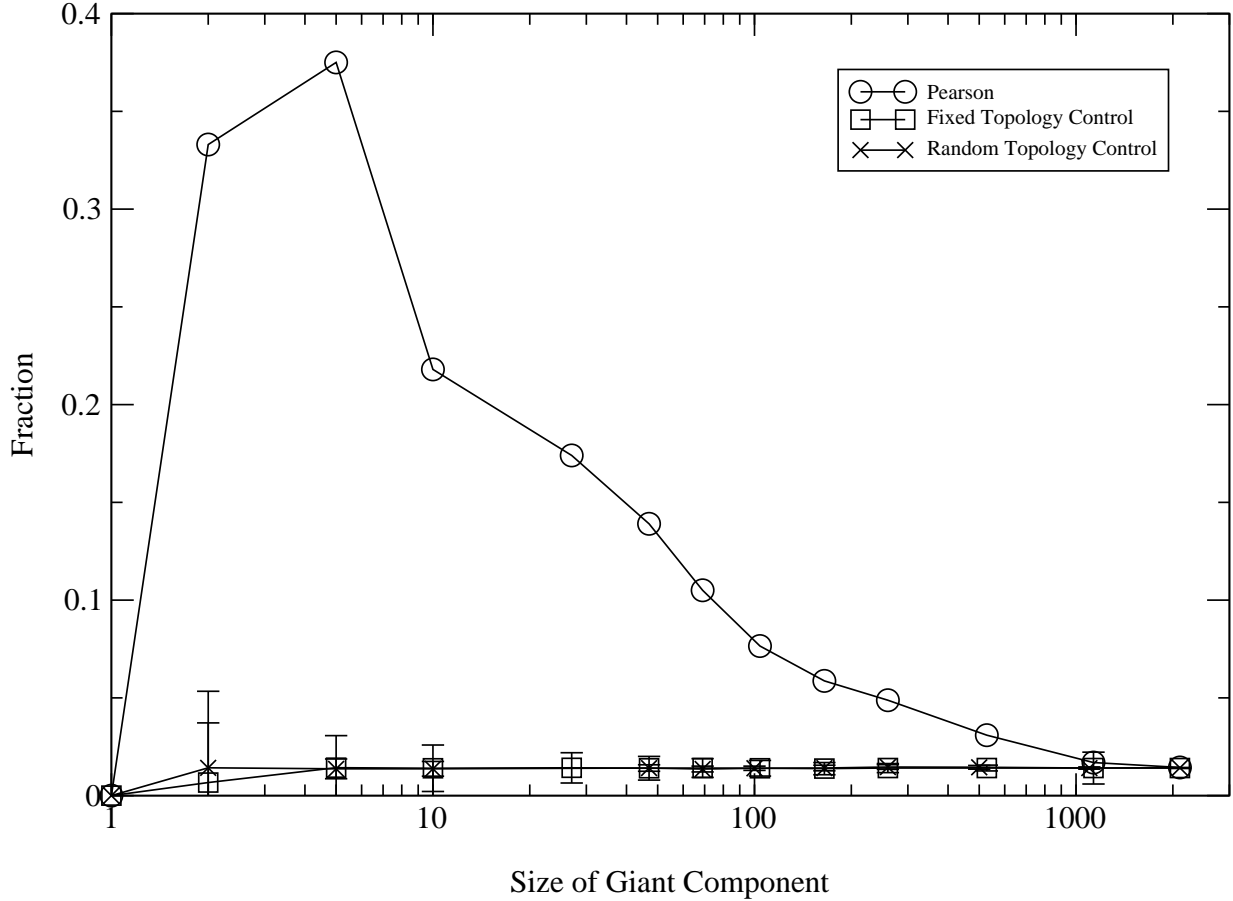}
}
\caption{\label{fig:probpair} 
The fraction of gene pairs clustered together and 
annotated with the same functional classification is 
shown for networks with varying values of the giant 
component.  Clusters were generated using the Pearson 
correlations of repeated microarray measurements (circles).
One random control was generated from randomized 
classifications on the fixed graph topology of the Pearson 
clustering (squares), and another was generated from 
graphs of variable topologies (crosses). } 
\end{figure}

To assess how well identically annotated genes were clustered, we
calculated the fraction of gene pairs that were in the same cluster and
annotated with the same functional classification, for different sizes of the giant component 
(Fig.~\ref{fig:probpair}).  
Specifically, any cluster of size $n$ has $(n)(n-1)/2$ total possible pairs. 
Thus a network of $m$ clusters with sizes $\{n_1, n_2, ... n_m\}$,
has a total number of possible pairings given by $\Sigma_{i=1}^{m}{(n_i)(n_i-1)/2}$.
Of these possible clustered pairings, a certain number will have matching functional
classification.
The calculated fraction is then the number of 
gene pairs with matching classification divided by the total number
of possible pairs.

This fraction increases with decreasing
giant component size until a peak value  
of 0.375, and then steeply drops when the size of the 
giant component reaches 1.  For permissive 
cut-offs and giant components exceeding more than
half the total number of genes, this fraction 
is near that of a random network, reflecting the loss
of information as the network becomes, trivially, 
completely connected.
Two types of random networks served as controls.
The first random control set was created by fixing the topology 
of the network, and randomizing the 
classification of the genes to calculate a complementary 
control probability.  
The second random control set was created by randomly linking
genes, in effect creating random topologies.
Comparing the pair matching fraction of the Pearson
data set to either null model shows substantial significance.
At the peak of the Pearson fraction curve in Fig.~\ref{fig:probpair},
the p-value is $2.12 \times 10^{-212}$
over the fixed topology control, and $8.08 \times 10^{-2226}$
over the random topology control. 

\begin{figure}[h]
\resizebox{\columnwidth}{!}{
\includegraphics{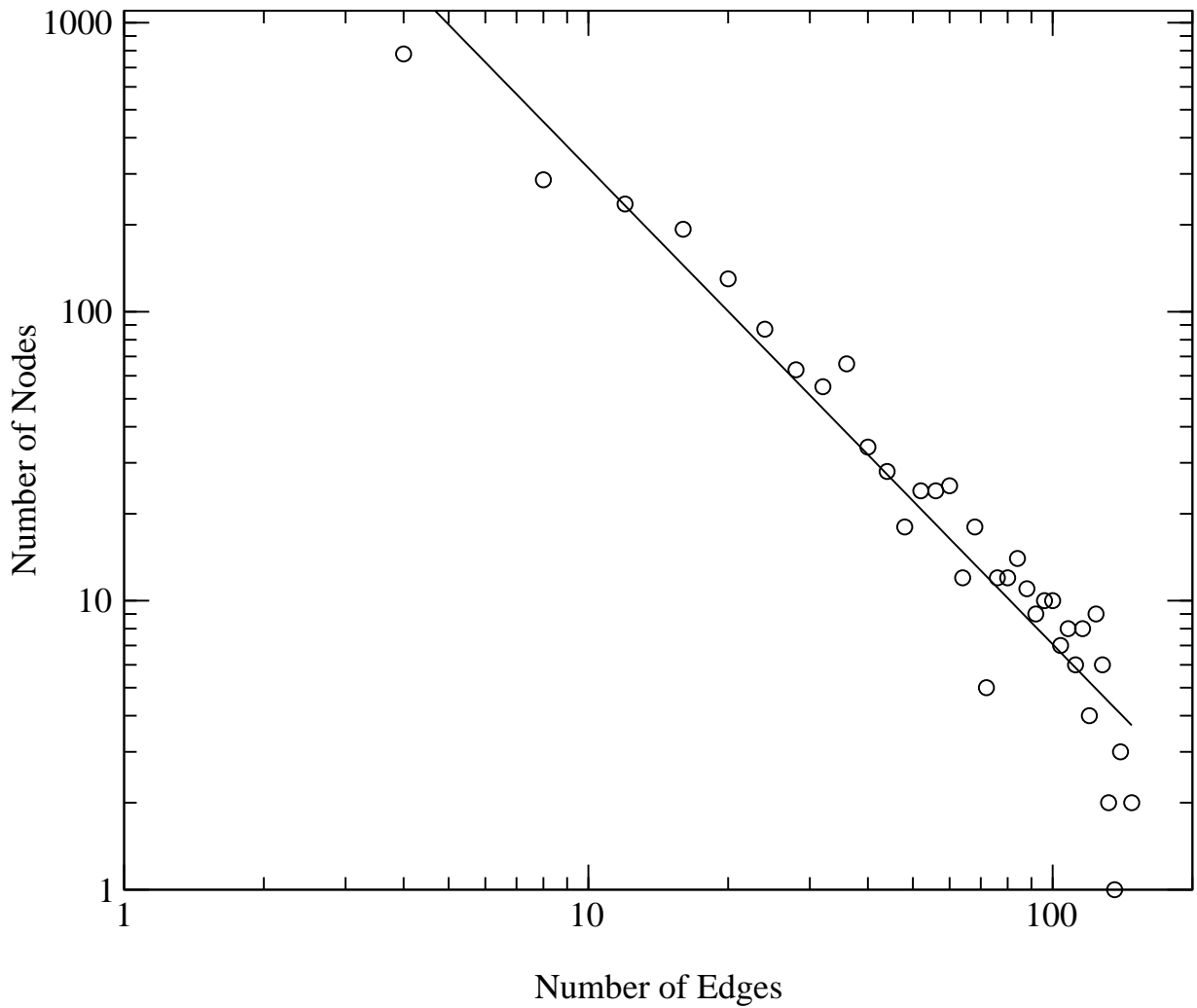}
}
\caption{\label{fig:powerlaw} 
Degree distribution of nodes in a network generated from 
Pearson correlation coefficents of microarray data.  Cut-off 
is chosen at the giant component transition (0.69).  The exponent of
the power law is -1.62.}
\end{figure}

At each cut-off, we also looked at a global property of the 
networks generated from noise data.  We calculated the 
degree distribution of the networks, which
is the distribution of the number of edges per node.
We show the degree distribution at the giant component
transition in Fig.~\ref{fig:powerlaw}.  The giant component
transition occurs at the cut-off (0.69) at which the size of the 
largest connected cluster is half the total number of nodes.

What is observed is a striking power-law.  
In a variety of other types of networks, 
power-laws are thought to reflect highly non-random assignment of links~\cite{albert_barabasi}.  
Such degree distributions have been identified previously in microarray-based gene clusterizations, but
have been restricted to the context of examining large-scale time-series changes in expression due to some drastic
change in the cell's overall state (e.g. diauxic shift or cell-division cycling)~\cite{bhan_dewey, bergmann_barkai}.
In contrast, we observe here that power-law behavior is 
visible in the mere fluctuations in expression across identically prepared cell samples.

The above analyses illustrate that the seeming randomness of mRNA expression fluctuations belies the presence of
highly structured, biologically significant patterns in the correlations of these fluctuations.  In what follows, we
propose a model which incorporates both deterministic and stochastic aspects into an explanation of why these
patterns might emerge.  Our goal is not to derive this model from any set of established, underlying equations known
to govern biological systems.  Rather, our purpose is to keep the model as simple and analytically tractable as
possible, and to let the question of its validity in describing gene networks be decided based on its empirical
success or failure in the laboratory.

\section{The Model}

We seek to model the stochastic dynamics of gene expression.  A natural starting point for this undertaking would seem to be the
description of expression in a single cell.  This would certainly be the most sensible approach if it were our intention to test
our model using data measured on a single-cell basis.  However, many high-throughput gene expression assays (such as mRNA
microarray experiments) are carried out on samples containing populations of cells.  Within these populations, it is certainly
reasonable to expect \textit{a priori} that there might be correlations in the noise experienced from one cell to another.  Thus, in order
to be as general as possible, we begin by assuming that we must describe some population of cells as a single, irreducible system,
although this does not rule out the possibility that the same simple model of gene expression might also be applied gainfully to
the study of single cells.

We begin by postulating, for the time being, that the total levels of all types of mRNA throughout the 
cell population fully characterize the state space of the system.  
This is obviously a gross over-simplification of real gene expression dynamics; a gene's activity can be regulated
post-transcriptionally through processes as diverse as alternative splicing, RNA interference, phosphorylation, and
proteolysis.  Furthermore, an mRNA's region of localization can in many cases be at least as important to its impact as
the amount of it that is transcribed, particular if it is found in some cells in the population, but not in others~\cite{ghaemmaghami_weissman, huh_oshea}!

Nevertheless, since our aim is to develop the simplest possible model of gene expression that will grant us some
predictive power, we proceed with our assumption despite these valid criticisms.  

Our next assumption is that, in the absence of noise, the expression levels (total mRNA levels) in the population of cells
would reach a stable, steady-state profile.  From a biological perspective, such a premise may at first seem totally
indefensible: a cell often passes through successive, cycling stages of growth during which aspects of its expression
profile must change dramatically. However, we may wish to restrict
our discussion for the moment to the modeling of growth-arrested cells which, for the most part, lack cycling degrees of
freedom.  Furthermore, since we are modeling the expression dynamics for a population of cells, it may be the case that
individual members of this population are at differents points in the cell cycle, but that a steady state has been
achieved, on average, for the population as a whole.

We also require, still in the absence of noise, that the rate at which the expression level of one gene changes
at a given time is some analytic
function of the current expression levels of all genes in the system.  Though it is reasonable to think that such a
view of gene expression is at least approximately valid for describing a network of genes affecting each other's rate
of production, it
is also entirely possible that it is factually incorrect. Current changes in expression may, for example, depend in a
discontinuous fashion
on whether certain levels are above or below some threshold~\cite{vilar_leibler, hasty_collins},
or else they may depend more strongly on the expression
profile at some finite time in the past than on the current profile.  The sensibility of this analyticity requirement
can only be judged by how accurate the quantitative predictions of the model prove to be in the laboratory. 

Finally, we assume that the rate of expression of each gene is subject to random noise which is uncorrelated from gene
to gene.  Rather than attempting to
tie this noise directly to any particular source, we leave its origin deliberately vague.  Our
aim is solely to introduce a stochastic aspect to the dynamics of our simple phenomenological model, without 
attempting to explain things in terms of underlying biological mechanisms.

The next step is to formalize the simple model that is implicit in these assumptions. 

In a system of $N$ genes, let the element $x_{i}$ in the vector $\mathbf{x}=
[x_{1},\ldots,x_{N}]$ describe the deviation of the expression level of gene $i$ from its steady-state level.  In this case, we have
\begin{equation}
\frac{dx_{i}}{dt}=F_{i}(\mathbf{x})
\end{equation}
where the steady-state condition is expressed as $\mathbf{F}(0)=0$.  If we define
\begin{equation}
K_{ij}=-\frac{\partial F_{i}}{\partial x_{j}}\bigg|_{\mathbf{x}=0}
\end{equation}
then, for small deviations from the steady-state in the absence of noise, we obtain
\begin{equation}
\frac{d\mathbf{x}}{dt} = -\mathbf{Kx}\label{linear}
\end{equation}
Demanding that the steady-state (SS) be stable is clearly equivalent to 
requiring the real parts of all eigenvalues of $\mathbf{K}$ to be
positive so that all trajectories eventually decay to the SS.  

Thus far, our model has followed the same course of previous theoretical work on linear response experiments in gene
networks
\cite{collins_gardner}.  At this point, however, we move in a new direction by introducing
noise to the system, which can be achieved by simply adding a randomly distributed quantity $f_{i}(t)$ to the right hand
side of Eq.~\ref{linear}:
\begin{equation}
\frac{d\mathbf{x}}{dt} = -\mathbf{Kx} +\mathbf{f}(t)\label{basic}
\end{equation}
The above $N$-dimensional first order heterogeneous equation was 
first investigated by Ornstein and Zernicke~\cite{ornstein_zernike}
as a class of Langevin models.  Here we reinterpret it to be the 
equation of motion for gene expression dynamics.
In another work, a Langevin model has been used
in the single gene context to explore
dynamics of transcription and translation ~\cite{ozbudak_vanoudenaarden}.
However, our work is contrasted by the fact that 
we are specifically interested in a many-gene network 
and the effects of noise on this network.

The simplest, non-trivial type of noise our system 
can experience is so-called uncorrelated white noise, which is
defined through the following:

\begin{eqnarray}
\langle \mathbf{f}(t) \rangle & = & 0 \\
\langle \mathbf{f}(t)\mathbf{f}^T(t') \rangle & = & 2T\,\mathbf{I}\delta(t - t')
\end{eqnarray}

The matrix $\mathbf{I}$ is the identity matrix. 
$T$ is some fictitious parameter that governs the 
global noise strength.  Here the $\langle...\rangle$ 
average is an ensemble average over all possible noise 
trajectories of the system.

Implicit in the noise distribution described above is the 
assumption that the magnitude of noise experienced by each
gene is the same.  More generally, it is possible that the 
noise correlation is fixed by some general diagonal matrix, 
$\langle \mathbf{f}(t)\mathbf{f}(t')^T \rangle = 2T\mathbf{\Lambda}\delta(t - t')$.
Thus, when comparing the predictions of the model to experiment, 
it might be necessary to treat the different elements
of $\mathbf{\Lambda}$ as free, unknown parameters.
The differences in the magnitude of noise may have a 
simple biological underpinning.  For example, large relative 
fluctuations can arise from low expression levels.
For this study, however, we consider the analytically
simpler case of uniform noise magnitude.

The formal solution to the equation of motion is given by

\begin{equation}
[\mathbf{x}(t)] = \int_0^t dt'\,[\exp(-\mathbf{K}t')]\mathbf{f}(t') +
[\exp(-\mathbf{K}t)]\mathbf{x}_0
\label{solution}
\end{equation}

The solution is a functional that depends on the noise trajectory ${\mathbf{f}(t)}$.  Since any one 
specific realization of the noise is uninteresting, we would like to 
know the statistical behavior of the system over
all possible realizations.  In other words, we would like to 
obtain the \textit{propagator} of the system, which gives 
the probability that the cell population attains a state 
$\mathbf{x}(t)$, given that it began in some initial state 
$\mathbf{x}_0$ at time zero. 

The propagator of Eq.~\ref{solution} is 
given by averaging a $\delta$-function of $[\mathbf{x}(t)]$ 
over all possible time trajectories of noise 
${\mathbf{f}(t)}$.

\begin{eqnarray}
G(\mathbf{x},\mathbf{x}_0;t) & = & \langle \delta(\mathbf{x} - [\mathbf{x}]) \rangle \label{defnprop} \\
\langle ... \rangle & = & \int \mathcal{D}\mathbf{f}(t)\, ...\, \exp\biggl[-\int_0^t dt'\, \mathbf{f}^2(t')\biggr]
\label{pathavg}
\end{eqnarray}

Substituting Eq.~\ref{solution} into Eq.~\ref{defnprop} 
and performing the Gaussian path integral in Eq.~\ref{pathavg} 
yields the propagator. 

\begin{equation}
G(\mathbf{x},\mathbf{x}^0;t) =
\biggl({ \frac{\pi}{\mbox{det}\,\mathbf{U} } }\biggr) ^\frac{N}{2} 
\exp[(-(\mathbf{x}^T - \mathbf{x}_0^T\exp(-\mathbf{K}^Tt))\mathbf{U}(t)(\mathbf{x} -
\exp(-\mathbf{K}t)\mathbf{x}_0))] \label{finalg}
\end{equation}

The propagator is complicated by the argument $\mathbf{U}(t)$
of the exponential.  The $\mathbf{U}(t)$ matrix of the
argument is only expressible in terms of an integral
and is not easily simplifiable except for special cases of $\mathbf{K}$.

\begin{equation}
\mathbf{U}(t) = \biggl[\biggl(\frac{1}{2T}\biggr)\,\int_0^t\,dt'\, \exp(-\mathbf{K}^Tt')\exp(-\mathbf{K}t')\biggr]^{-1}
\label{ureln}
\end{equation}

For finite time, the inverse of the $\mathbf{U}(t)$ matrix 
gives the equal time correlators of the system.

\begin{equation}
\mathbf{U}^{-1}(t) = \langle \mathbf{x}(t) \mathbf{x}^T(t) \rangle
- \langle \mathbf{x}(t) \rangle \langle \mathbf{x}^T(t) \rangle
\end{equation}

While time-dependent pheneomena are very interesting, 
the experimental measurement of noise-averaged, time-dependent quantities
is likely to be a technically challenging task.
We therefore focus our attention for the present on the simpler case of taking
repeated, uncorrelated expression measurements from identically prepared cell samples.
This is equivalent to probing the system in its 
\textit{equilibrium} limit, when traces
of the initial conditions are lost.  This condition can be 
realized by taking the infinite-time limit of 
Eq.~\ref{finalg} and~\ref{ureln}.

\begin{eqnarray}
G_{eq}(\mathbf{x}) & = & 
\biggl({ \frac{\pi}{\mbox{det}\,\mathbf{U}_{eq} } }\biggr) ^\frac{N}{2}
\exp{(-\mathbf{x}^T\,\mathbf{U}_{eq}\,\mathbf{x}}) \\ \label{eqprop}
\mathbf{U}_{eq} & = &\biggl[ \biggl(\frac{1}{2T}\biggr)\,\int_0^{\infty}\,dt'\,\exp(-\mathbf{K}^Tt')\exp(-\mathbf{K}t')\biggr]^{-1} \label{Ueq}
\end{eqnarray}

Thus, we obtain an expression relating the covariance of fluctuations in expression measurements to the underlying
matrix of interactions between the different genes in the network: 
\begin{equation}
\mathbf{C}\equiv\langle\mathbf{x}~\mathbf{x}^{T}\rangle -\langle\mathbf{x}\rangle\langle\mathbf{x}^{T}\rangle \propto
\int _{0}^{\infty}dt
\exp(-\mathbf{K}^Tt) \exp(-\mathbf{K}t)\label{covar}
\end{equation}

\section{Results and Discussion}
Eq.~\ref{covar} is striking because it describes an exact relationship that must exist between the
rate constants of a deterministic gene network near a steady-state and the expression fluctuations
that will result if such a system is driven by random noise ($\mathbf{C}$).  Interestingly, the relationship implied
by our model
provides us with a means to understand how biological information contained in the underlying interactions between genes
($\mathbf{K}$) can be preserved to some extent in the covariance matrix  ($\mathbf{C}$) that we examined in Fig.~\ref{fig:probpair}. 
For the sake of being most illustrative, let us assume for the moment that $\mathbf{K}$ is a symmetric matrix that can
be expressed as the difference of the identity matrix $\mathbf{I}$ and a symmetric off-diagonal matrix $\mathbf{k}$, all
of whose
eigenvalues are less than unity in magnitude.   In this special case, Eq.\ref{covar} can be simplified, yielding the simple result
\begin{equation}
C_{ij}\propto
(\mathbf{K}^{-1})_{ij}=(\mathbf{I}-\mathbf{k})^{-1}_{ij}=\delta_{ij}+k_{ij}+\sum_{l}k_{il}k_{lj}+\sum_{lm}k_{il}k_{lm}k_{mj}+\ldots
\label{pwseries}
\end{equation}
In other words, the covariance of fluctuations in genes $i$ and $j$ is obtained by summing over all of the ways in
which the two genes are indirectly coupled to each other across the network of interactions.  Eq.~\ref{pwseries}
indicates that whatever biologically meaningful structure might originally be represented in
$\mathbf{k}$, it is likely to be preserved, and possibly even reinforced, in the subsequent correlation fluctuations
described in $\mathbf{C}$, since it is likely that genes of similar function will be more
highly connected to each other through indirect routes across the network. 

Even more significant than the qualitative result discussed above, however, is the rigorously testable quantitative
prediction that is also implicit in Eq.\ref{covar}. Obviously, it is possible to measure $\mathbf{C}$.  If we were also
able to determine all of the linear rate constants contained in the matrix $\mathbf{K}$ by some independent means, then
we would be able to calculate $\mathbf{C}$ using the model and then compare the predicted covariances to the
experimentally measured ones.

A straight-forward method of determining $\mathbf{K}$ has already been proposed and implemented on a small scale
\cite{collins_gardner}.  The matrix elements of $\mathbf{K}$ can be calculated by measuring the linear response of 
the system's steady state to a small, constant, exogenous increase in the rate of production of some gene in the network.
More formally, this amounts to introducing a ``source term'' into (\ref{basic}):  

\begin{equation}
\frac{d \mathbf{x} }{dt} = -\mathbf{K} \mathbf{x} +\mathbf{h}+\mathbf{f}
\end{equation}
which gives
\begin{equation}
\frac{d( \mathbf{x}-\mathbf{K}^{-1}\mathbf{h}) }{dt} = -\mathbf{K} ( \mathbf{x}-\mathbf{K}^{-1}\mathbf{h})+\mathbf{f} 
\end{equation}
By measuring the SS shift 
of every gene $x_i$ in response to a source 
$h_j$ pointing in the 
``direction'' of the $j^{th}$ gene, we obtain one
column of the inverse of $\mathbf{K}$:
\begin{equation}
\frac{\partial \langle x_i \rangle_{eq}}{\partial h_j} = (\mathbf{K}^{-1})_{ij}
\end{equation}
It should be possible to introduce such a source experimentally in a number of ways, such as by
insertion of  an inducible plasmid in \textit{S. cerevisiae}, 
or by effecting low, quantitative levels of RNAi 
in \textit{Drosophila} cell culture.  The validation of the model's equilibrium predictions using such experimental
approaches would have remarkable implications for our understanding of the dynamics of these systems over shorter
times scales.

Another intriguing possibility is that one might use experimental measurements of $\mathbf{C}$ in
order to extract information about the linear response coefficients $\mathbf{K}$.  In the special case examined above,
we were able to derive a closed-form expression for $\mathbf{K}$ in terms of $\mathbf{C}$.  In the more general 
and realistic case of
a non-symmetric $\mathbf{K}$, however, the noise erases some of the information we need in order to completely
reconstruct $\mathbf{K}$ from $\mathbf{C}$.

To see this most clearly, it is helpful to consider the cases where either the symmetric or anti-symmetric part of $\mathbf{K}$
is small.  In either instance, we can then Fourier transform our expression for $\mathbf{C}$ and obtain the
approximate result
\begin{equation}
\mathbf{C}^{-2}\propto \mathbf{K}\mathbf{K}^{T}\label{phase} 
\end{equation}
If we think of $\mathbf{K}$ as the an $N$-dimensional analog of a complex number, then we can see that while it is
possible to recover the ``magnitude'' of $\mathbf{K}$  using $\mathbf{C}$, the noise has caused us to lose information
about the ``phase''.  Put another way, if we were able to find one matrix which satisfied (\ref{phase}), we could
generate another by applying any $N$-dimensional rotation to the rows of $\mathbf{K}$ 

It is conceivable that a well-designed set of experiments might provide us with the complementary information
necessary to recover $\mathbf{K}$ from $\mathbf{C}$.  This raises the exciting possibility that the dynamical
coefficients of large gene networks might be obtainable without the need for conducting large numbers of painstaking
linear response measurements.  Alternatively, it may be fruitful to follow the example of Gardner et. al.
\cite{collins_gardner} and make simplifying assumptions about the underlying structure of $\mathbf{K}$ which would
make its construction from $\mathbf{C}$ into an overdetermined problem.

One point that remains to be addressed is the issue of sampling.  In any statistical system with
multiple
variates, the covariance matrix obtained from a finite sampling of the distribution is singular, and therefore not
invertible, at least until the sample size exceeds the number of variates.  This immediately implies that good sampling
can only be achieved with microarrays for relatively small networks because of the costliness of the technology. 

This stumbling block might be avoided, however, if we were to use \textit{protein} as our proxy for gene expression
instead of mRNA (This substitution is a perfectly acceptable one as far as the model is concerned, since it was
constructed to be a rough, general model of gene expression).  In this case, it would be possible to prepare cell
lines with pairs of proteins labeled with different fluorescent reporters~\cite{huh_oshea}. 
Through the use of FACS methods, one could therefore measure the expression covariance of genes using thousands of independent samples.

Finally, it should be noted that, though we have focused on a linear network model here, the theoretical methods we have
employed here can be easily generalized to incorporate non-linear effects.  Using a related path integral technique, it
should be possible to perturbatively account for higher-order couplings between different genes in the network, as
well as more sophisticated noise distributions.  Such an undertaking, however, would be much more involved, and should
almost certainly be postponed until the limitations of the linear model are probed in experiment.
 
In this study, we presented evidence for the presence of ample amounts of biological information in the expression
fluctuations of unperturbed
genetic networks, and we introduced a dynamical model
of gene expression that seeks to explain the informational content of these fluctuations.  
We believe our model to be the first to incorporate the intrinsic
noisiness of gene expression into an exactly solvable dynamical description 
of the web of genetic interactions that exists in a cell. 
We used the model to derive a non-trivial prediction about the relationship between the dynamical linear response
coefficients of the network near a steady-state and the correlations which must exist among steady-state fluctuations in
the expression of different genes over time. 
Given enough experimental data, it is not inconceivable that at some future time, researchers
will be able to model genetic networks exactly with detailed simulations on powerful computers.  The spirit of this work 
lies at the other end of the spectrum; we are hopeful that the
simple, coarse-grained model presented here incorporates enough of the salient features of
genetic networks that it may provide some analytical insight into their fundamental nature in addition to being a useful
tool for predicting their behavior.

\small{We thank Boris Shakhnovich for useful discussions.  We also thank Siraj Ali 
and Saeed Tavazoie for pointing us in the direction of the Hughes et al. data set.  
This work was supported by the National Institutes of Health.}

\end{document}